\documentclass[12pt]{article}

\usepackage[utf8]{inputenc}
\usepackage[T1]{fontenc}
\usepackage{graphicx}
\usepackage{amsmath,amssymb}
\usepackage{booktabs}
\usepackage{hyperref}
\usepackage{cleveref}
\usepackage{subcaption}
\usepackage{wrapfig}
\usepackage{url}
\usepackage[margin=1in]{geometry}
\usepackage{listings}
\usepackage{xcolor}

\title{LoRa and LoRaWAN simulator-cum-emulator with CAD and capture effect in Python}
\author{Matthijs Reyers, Niels Hokke, R.R.\ Venkatesha Prasad}
\date{May 2026}

\begin{document}

\maketitle



\begin{abstract}


Existing LoRaWAN/LoRa simulators consist of large, complicated C++ codebases and often do not support all device classes.
This paper presents the design of a simple to use, Python-based discrete-event simulator that addresses these gaps while also introducing a novel method for evaluating real device firmware in the simulator.
The simulator is built on a custom \texttt{asyncio}-based simulation kernel, a three-phase packet delivery model that reproduces the capture effect, a full LoRaWAN~1.0.4 stack, and a containerized firmware system that cross-compiles real STM32 C firmware and redirects HAL calls into the simulator via CFFI.
The simulator is distributed as a Python package via Github (\url{https://github.com/MatthijsReyers/lora-simulator}) and requires no external simulation framework or dependencies.
\end{abstract}


\section{Introduction} \label{sec:introduction}

LoRa is a spread-spectrum radio technology developed by Semtech that enables low-power, long-range communication for Internet of Things (IoT) devices~\cite{SemtechAcquiresCycleo}.
The most widely adopted MAC-layer protocol built on top of LoRa is LoRaWAN, which defines device classes (A, B, C), over-the-air activation (OTAA), and a full network architecture with gateways and network servers~\cite{haxhibeqiri2018survey}.
While LoRaWAN is the default choice for most deployments, some companies and research groups opt to develop custom protocols directly on top of the LoRa physical layer---for example to reduce energy consumption, to support application-specific scheduling, or to implement firmware update over the air (FUOTA) strategies that do not fit within the constraints of standard LoRaWAN.

Simulation plays a critical role in developing and evaluating such protocols, particularly for scenarios that are difficult or expensive to test with physical hardware at scale.
Several LoRa simulators exist (Section~\ref{sec:related-work}), but none simultaneously support LoRaWAN class A, class B, and class C devices with capture effect and CAD modeling\cite{kouvelas2018employing}, protocol-agnostic custom MAC layers, firmware-in-the-loop emulation of real embedded C code, and per-device energy tracking, without requiring users to extend a complex C++ simulation framework.

This paper presents the design of an open-source, Python-based simulator that addresses these gaps.
The key contributions are a three-phase packet delivery model that faithfully reproduces the LoRa capture effect based on the interference model of Haxhibeqiri et al.~\cite{haxhibeqiri2017lora}, and a containerized firmware system that cross-compiles real STM32 C firmware for the host architecture and redirects HAL calls into the simulator via CFFI without requiring a full hardware emulator.

\section{Related work} \label{sec:related-work}

Several LoRa simulation environments have been proposed in the literature.
This section summarizes the most prominent tools and highlights the gaps that motivate the present work.

\textbf{NS-3 LoRaWAN}~\cite{8422800} is a C++ module for the widely used NS-3 network simulator.
It provides a full LoRaWAN protocol stack with energy modeling and configurable path loss, but supports only Class~A devices, has no multicast or FUOTA capabilities, and cannot execute real firmware.

\textbf{FLoRa}~\cite{ComparisonLoraSimEnv} is built on OMNeT++ with the INET framework.
It supports ADR and provides a graphical interface, but is similarly limited to Class~A and does not model the capture effect or support firmware-in-the-loop testing.

\textbf{LoRaSim}~\cite{haxhibeqiri2017lora} is a lightweight Python simulator built on SimPy.
It models the LoRa physical layer including the capture effect and supports multi-gateway deployments, but does not implement any LoRaWAN MAC-layer functionality, is written in Python~2, and has been inactive since 2017.

\textbf{LoRa-MAB}~\cite{ComparisonLoraSimEnv} focuses on decentralized resource allocation using multi-armed bandit algorithms and is not designed for general-purpose protocol evaluation.

Table~\ref{tab:simulator-comparison} compares these simulators against the five requirements identified in Section~\ref{sec:introduction}.
No existing tool satisfies all requirements, which motivates the development of the simulator presented in this paper.

\begin{table}[htbp]
\centering
\caption{Comparison of existing LoRa simulators against the requirements identified in this paper.
\checkmark\ = supported, $\sim$ = partial, \texttimes\ = not supported.}
\label{tab:simulator-comparison}
\small
\begin{tabular}{@{}lccccc@{}}
\toprule
\textbf{Requirement} & \textbf{NS-3} & \textbf{FLoRa} & \textbf{LoRaSim} & \textbf{LoRa-MAB} & \textbf{Ours} \\
\midrule
All device classes + multicast  & \texttimes & \texttimes & \texttimes & \texttimes & \checkmark \\
Capture effect + CAD            & \checkmark & \texttimes & \checkmark & \checkmark & \checkmark \\
Protocol agnosticism            & \texttimes & \texttimes & $\sim$     & \texttimes & \checkmark \\
Firmware-in-the-loop            & \texttimes & \texttimes & \texttimes & \texttimes & \checkmark \\
Energy model                    & \checkmark & \checkmark & \texttimes & \texttimes & \checkmark \\
\midrule
\textbf{Score}                  & 2/5        & 1/5        & 1.5/5      & 1/5        & 5/5 \\
\bottomrule
\end{tabular}
\end{table}

\section{Design goals} \label{sec:design-goals}

\subsection{Ease of use}
Existing LoRa simulators such as NS-3 or OMNeT++ are written in C++ and support many different radio technologies and protocols.
While this makes them very powerful, it creates a high barrier to entry for researchers who only need to simulate custom LoRa protocols, because they must first become familiar with large codebases and C++ programming.
With a purpose-built simulator in Python---a widely used language~\cite{StatistaMostUsedLanguages}---the process of setting up and running a simulated LoRa network is much simpler.
Simulations are defined as ordinary Python scripts that import the simulator as a standard library, meaning a simple simulation can be a single file of fewer than 100 lines of code.

\subsection{Modular and extensible}
No simulator can anticipate every experiment a researcher might want to run.
The simulator is therefore designed so that individual components---path-loss models, capture effect detection, MAC-layer protocols---are modular and can easily be replaced or extended.
This loose coupling is also beneficial for testability: components with well-defined interfaces can be unit-tested and validated in isolation, giving confidence that any discrepancy found in simulation reflects a protocol-level issue rather than a simulator artifact.

\subsection{Real firmware integration}\label{sec:design-goals:real-firmware}
A fundamental limitation of behavioral simulators is that the model of the device under test is written separately from the firmware that will actually run on the hardware.
Bugs or subtle timing dependencies in the real firmware will not surface in simulation, and results may diverge from hardware measurements in ways that are difficult to diagnose.
To close this gap, the simulator should be capable of running actual C/C++ firmware source code directly, without requiring a full hardware emulator.

\subsection{Protocol agnosticism}\label{sec:design-goals:protocol-agnostic}
Existing LoRa simulators are tightly coupled to the standard LoRaWAN Class~A device model.
However, many use cases---particularly FUOTA---require Class~C continuous receive windows, multicast groups, or entirely custom MAC-layer protocols.
The simulator must therefore be protocol-agnostic at its core: the LoRaWAN stack should be an optional layer on top of the physical layer, and users must be able to implement and test arbitrary protocols without modifying the simulator itself.

\section{High-level architecture} \label{sec:sim-architecture}

The simulator is written in Python and built on a custom discrete-event simulation kernel that uses the \texttt{asyncio} standard library as its foundation.
Unlike LoRa simulators that use SimPy~\cite{haxhibeqiri2017lora}, the simulator uses native \texttt{asyncio} tasks and events as its scheduling primitives.
This design decision was motivated by two concerns: first, \texttt{asyncio} is part of the Python standard library and does not add an external dependency; second, \texttt{asyncio}'s task model maps cleanly to the concurrent nature of a wireless sensor network where each node runs its own independent control loop.

The architecture is organized into four layers, each building upon the one below it:

\begin{enumerate}
    \item \textbf{Simulation kernel} --- manages virtual time, task scheduling, and inter-task synchronization.
    \item \textbf{LoRa physical layer} --- models the wireless medium, packet propagation, path loss, collision detection, and the capture effect.
    \item \textbf{Protocol stack} --- implements LoRaWAN~1.0.4 (framing, cryptography, MAC commands, OTAA, device classes) and provides hooks for custom protocols.
    \item \textbf{Containerized firmware} --- enables real C firmware to run inside the simulator via cross-compilation and CFFI-based HAL interception.
\end{enumerate}

\begin{figure}[htb]
    \centering
    \includegraphics[width=\textwidth]{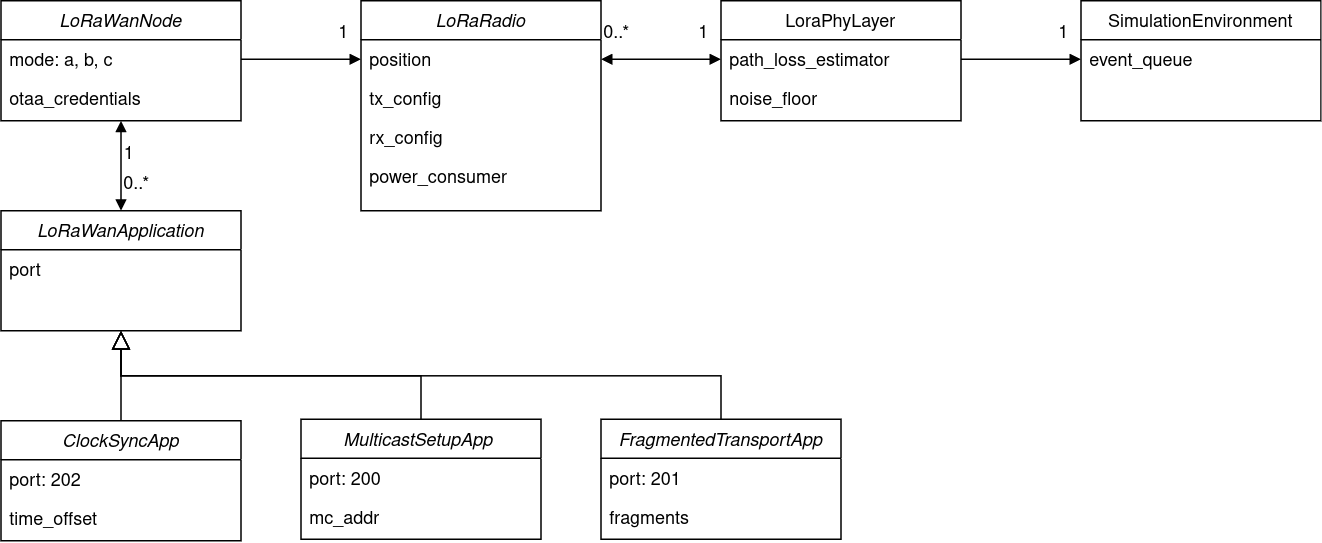}
    \caption{Simplified UML class diagram of the simulator's core components. \texttt{LoraRadio} instances register with the singleton \texttt{LoraPhyLayer} and are owned by a \texttt{LoRaWanNode}. Application-layer logic is decoupled via the abstract \texttt{LoRaWanApplication} class, of which three standard implementations are provided.}
    \label{fig:simulator-classes}
\end{figure}


Figure~\ref{fig:simulator-classes} shows the relationships between the simulator's core classes.
The \texttt{SimulationEnvironment} singleton manages the event queue and virtual clock; the \texttt{LoraPhyLayer} singleton models the shared wireless medium and holds references to all registered \texttt{LoraRadio} instances.
Each radio encapsulates its position, TX/RX configuration, and a \texttt{PowerConsumer} for energy accounting.
The \texttt{LoRaWanNode} class---which owns one radio and zero or more \texttt{LoRaWanApplication} instances---implements the full LoRaWAN protocol stack, but is intentionally \emph{not} a required component: it serves primarily as a reference implementation showing how to build a protocol on top of the radio API.
Users developing custom LoRa protocols can instantiate a \texttt{LoraRadio} directly and implement their own node logic without inheriting from or depending on any LoRaWAN class. 

\subsection{Simulation kernel}\label{subsec:sim-kernel}

At the core of the simulator is the \texttt{Simulation\-Environment} (\texttt{simulator/environment.py}), a singleton object imported by all other modules.
Time is discretized into configurable ticks (default: 1\,\textmu s) and only advances when \emph{no task is actively executing}.
This is enforced through a timer-lock counter: each running task increments the counter upon waking and decrements it upon yielding.
When the counter reaches zero, the kernel inspects a priority queue of scheduled wakeup events (\texttt{WakeUpQueue}) and jumps directly to the earliest pending event, skipping any intermediate empty ticks.
This event-skipping approach makes the simulator efficient even for long simulated durations with sparse activity.

The kernel exposes a small API that all higher layers use:
\texttt{sleep(duration)} suspends a task for a simulated duration,
\texttt{sleep\_until(timestamp)} suspends until an absolute simulation time,
\texttt{create\_task()} registers a root-level coroutine before the simulation starts,
and \texttt{start\_child\_task()} spawns a new task during the simulation.
A simulation-aware generic \texttt{Queue[T]} class is also provided, which uses scheduled \texttt{asyncio.Event} objects so that consumers block in simulated time rather than wall-clock time.

\subsection{Shared wireless medium}\label{subsec:sim-shared-medium}

The LoRa physical layer is modeled as a single shared medium represented by the \texttt{LoraPhyLayer} singleton (\texttt{simulator/lora/phy\_layer.py}).
All radio instances register themselves with this singleton upon creation.
When a radio transmits a packet, the PHY layer delivers a deep copy of the packet to every other registered radio, with each copy annotated with the receiver's location so that per-link path loss can be computed.

Packet delivery is split into three phases to enable realistic collision-window modeling:
\begin{enumerate}
    \item \textbf{Receive start} --- fired immediately when the transmission begins; each radio checks whether the incoming packet matches its current configuration (spreading factor, bandwidth, code rate, IQ inversion).
    \item \textbf{Preamble received} --- fired after the preamble airtime has elapsed; at this point the radio decides whether it can lock onto the packet or whether interference from an overlapping transmission prevents demodulation.
    \item \textbf{Receive end} --- fired after the full payload airtime; the packet is delivered to the radio's receive queue if no fatal collision occurred during the payload.
\end{enumerate}

This three-phase model allows the simulator to reproduce phenomena that single-phase collision models cannot, such as the capture effect (Section~\ref{subsec:sim-capture}), where a stronger packet arriving during another packet's preamble can still be received successfully.


\subsection{Protocol and application layer}\label{subsec:sim-protocol-app}

On top of the PHY layer, the simulator provides a complete LoRaWAN~1.0.4 protocol stack comprising a \texttt{LoRaWanDevice} (\texttt{simulator/lorawan/device.py}), a \texttt{LoRaWanGateway} (\texttt{simulator/lorawan/gateway.py}), and a \texttt{NetworkServer} (\texttt{simulator/lorawan/network\_server.py}).
The device supports both ABP and OTAA activation, Class~A and Class~C receive windows, multicast group membership, and a full set of MAC commands.
All cryptographic operations (AES-128 CMAC for MIC computation, CTR-mode payload encryption, ECB-mode join accept encryption, and session key derivation) are implemented in \texttt{simulator/lorawan/crypto.py} using the \texttt{cryptography} Python library and follow the LoRaWAN specification.

Application-layer logic is decoupled from the protocol stack through an abstract \texttt{Application} base class (\texttt{simulator/lorawan/application.py}).
Each application registers itself on a specific FPort (1--223); when a downlink or uplink arrives on that port, the corresponding application handler is invoked with the decrypted payload.
This design allows new application-layer protocols---including FUOTA-related protocols---to be added and tested without modifying the LoRaWAN stack itself.

Because the LoRaWAN stack communicates with the radio exclusively through the \texttt{LoraRadio} interface (\texttt{simulator/lora/radio.py}), users who need a custom protocol can bypass the LoRaWAN stack entirely and program their nodes directly against the radio API, which mirrors the STM32WL HAL function signatures.

\subsection{Language choice}\label{subsec:sim-language}

Python was chosen as the implementation language for several reasons.
Python is consistently ranked among the most widely used programming languages by professional developers worldwide~\cite{StatistaMostUsedLanguages}, which means that most researchers are likely to already be familiar with the language.
Its \texttt{asyncio} library provides native coroutine support that maps well to the concurrent task model of a sensor network.
Python's \texttt{cffi} library enables tight integration with C firmware code (Section~\ref{sec:sim-native-node}) without requiring a full hardware emulator.
The scientific Python ecosystem---particularly \texttt{pandas} and \texttt{numpy}---makes post-simulation data analysis straightforward (Section~\ref{sec:sim-logging}).
The main trade-off is execution speed: Python is slower than C++ simulators like NS-3 or OMNeT++.
However, for the network sizes targeted in this work (up to several hundred nodes), simulation runs complete in seconds to minutes, which is acceptable for iterative protocol development.

\section{Containerized firmware} \label{sec:sim-native-node}

In order to achieve the aforementioned integration of real firmware
(see Section~\ref{sec:design-goals:real-firmware}) within the simulator, we
use a system inspired by modern environment isolation systems like
Docker containers.
These containerization systems do not run a full virtual machine for
each container but instead intercept system calls made by the software
in the container and map them to a virtual kernel instead of the host
operating system's kernel.
This is advantageous because the computational overhead of intercepting
system calls is generally much lower than running a whole virtual machine.

The simulator similarly does not run a full STM32 hardware emulator
but instead cross-compiles the firmware source code for the host
machine's architecture,
while all HAL (Hardware Abstraction Layer) library implementation files
are replaced with shim implementations (\texttt{simulator/native\_node/stm32\_node.py}) that redirect hardware-specific
function calls to the simulator's event loop using CFFI (C Foreign Function Interface).
The HAL functions are essentially the system calls in our Docker container
analogy.
For example, \texttt{HAL\_Delay()} is redirected to the simulator's
virtual-time sleep function, and \texttt{HAL\_GetTick()} returns the
current simulation time in milliseconds instead of reading a hardware
timer register.

\begin{figure}[htb]
    \centering
    \begin{subfigure}[b]{0.45\textwidth}
        \includegraphics[width=\textwidth]{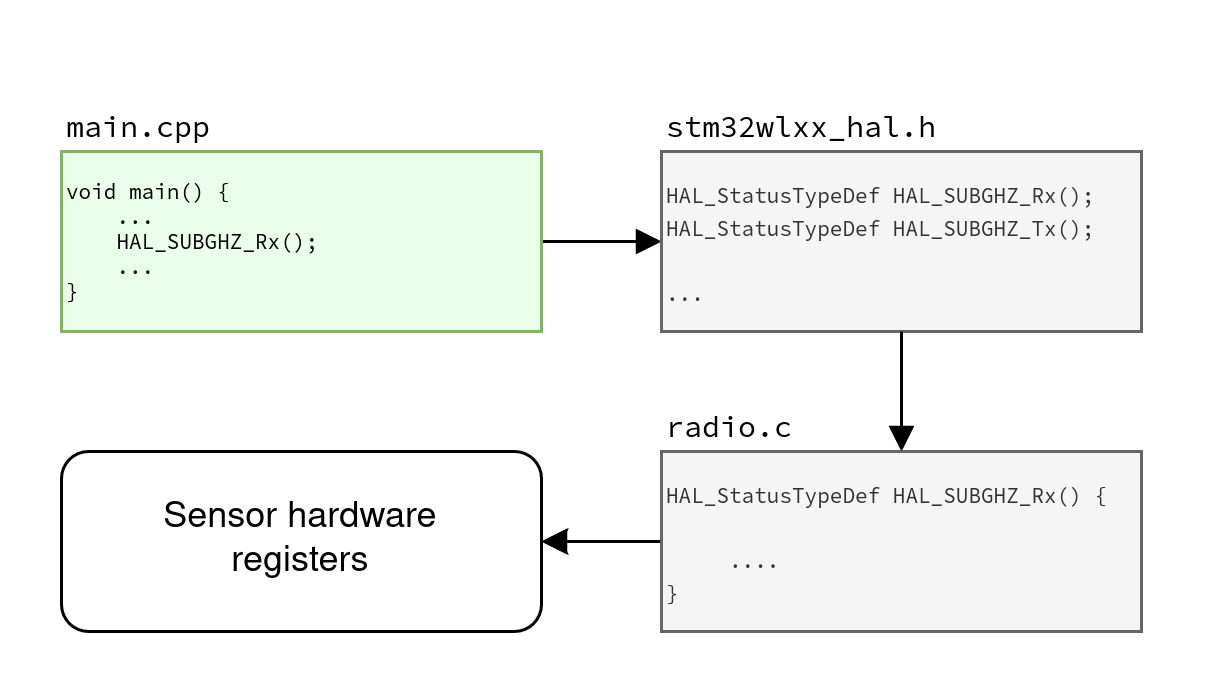}
        \caption{Normal firmware architecture}
        \label{fig:normal-firmware}
    \end{subfigure}
    \hfill
    \begin{subfigure}[b]{0.45\textwidth}
        \includegraphics[width=\textwidth]{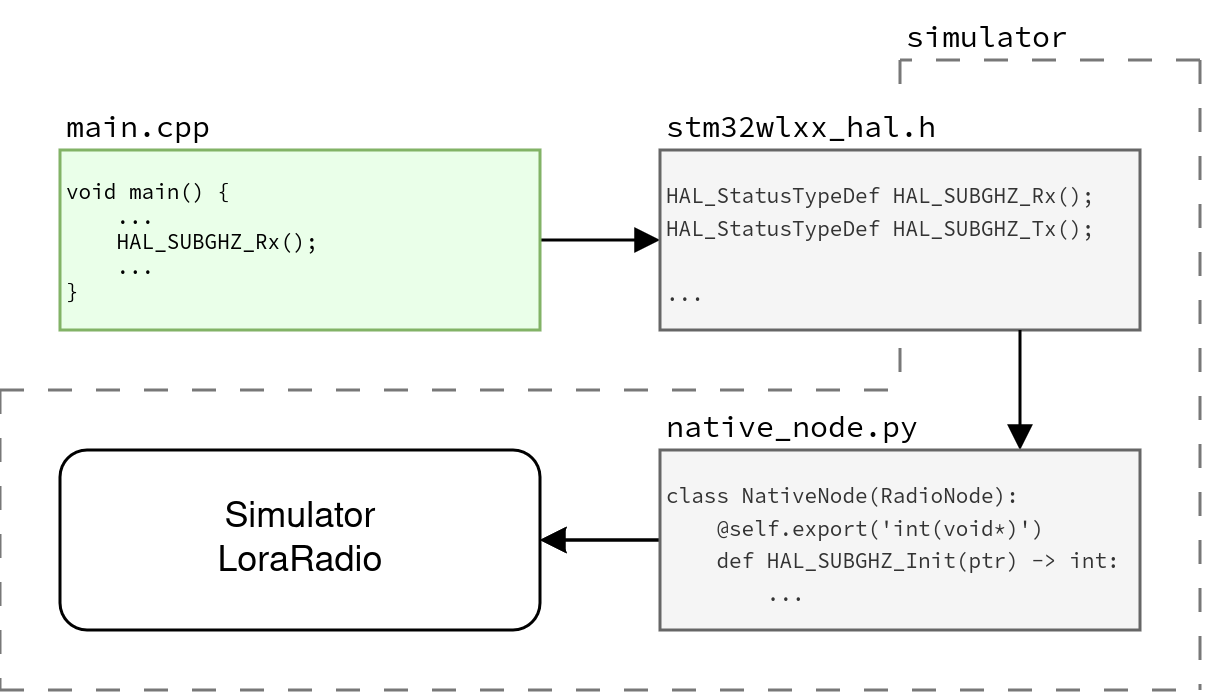}
        \caption{Simulated firmware architecture}
        \label{fig:simulated-firmware}
    \end{subfigure}
    \caption{Comparison of normal and simulated firmware architectures. In the containerized approach~(b), HAL implementation files are replaced with shims that redirect calls to the simulator via CFFI.}
    \label{fig:firmware-architectures}
\end{figure}

Figure~\ref{fig:firmware-architectures} shows the change in control
flow and compiler linking steps for the containerized firmware
(Figure~\ref{fig:simulated-firmware}) compared to firmware deployed on a real sensor
(Figure~\ref{fig:normal-firmware}).
The header files defining the data structures used by the HAL are kept
unchanged, as the goal is to ensure that no modifications have to
be made to the firmware code in order to run it in the simulator.
This is important since any simulator-specific modifications could cause
(or hide) bugs by changing the firmware's behavior.
Users of the simulator do have to provide their own shims for application-specific
hardware peripherals; for example, an accelerometer read out via SPI
requires a shim that returns synthetic or recorded sensor data.

A limitation of this system is that it can catch logic bugs in the firmware
code, but cannot guarantee detection of memory management problems
like buffer overflows or undefined behavior, because
these bugs may manifest differently on an
embedded ARM Cortex chip versus a desktop x86 CPU.
There is also the small chance of a compiler backend bug for a
specific architecture, though this is very unlikely with modern compilers.
This is a worthwhile trade-off, as cross-compilation and CFFI
save considerable complexity and compute power compared to running a full
virtualized STM32 chip.

\section{Logging \& analysis} \label{sec:sim-logging}

All modules of the simulator make use of the \texttt{logging} Python
standard library module.
This allows users to enable specific logging levels for individual parts
of the simulator independently; for example, one can enable verbose
\texttt{DEBUG}-level output from the PHY layer while keeping the rest
of the simulator silent.

These log messages are useful for real-time tracing but are not
well-suited for post-hoc analysis.
To support quantitative analysis, several key classes expose their
recorded data as pandas \texttt{DataFrame} objects after the simulation
ends.
Collecting data by appending rows to a \texttt{DataFrame} incrementally
during the simulation would be prohibitively slow for large runs.
Instead, each class accumulates records in an ordinary Python
\texttt{dict} of lists during the simulation and converts it to a
\texttt{DataFrame} exactly once when the simulation finishes, via a
callback registered on the \texttt{wait\_for\_sim\_end} hook.

Three classes expose results in this way:

\begin{itemize}
    \item \textbf{\texttt{LoraPhyLayer.packets\_log}}: one row per
        transmitted packet, with columns for spreading factor, bandwidth,
        code rate, airtime, transmit power, transmitter location, payload
        bytes, and the unique radio identifier of the sender.

    \item \textbf{\texttt{LoraRadio.packets\_log}}: one row per packet
        that arrived at a specific radio, with columns for received SNR,
        RSSI, and boolean flags indicating whether the packet suffered a
        collision, whether the radio missed the preamble,
        or whether reception was interrupted mid-packet.
        Because each node has its own \texttt{LoraRadio} instance, this
        table can be queried per node to compute metrics such
        as packet delivery ratio or average SNR.

    \item \textbf{\texttt{PowerConsumer.events}}: one row per power-state
        transition, recording the simulation time, the new power draw in
        watts, and the cumulative energy consumed in joules up to that
        point.
        Every \texttt{LoraRadio} owns a \texttt{PowerConsumer} instance
        (\texttt{simulator/power\_consumer.py}),
        so per-node energy budgets can be extracted directly.
\end{itemize}

Because all three tables are standard pandas \texttt{DataFrame} objects,
users can apply the full pandas and NumPy ecosystem for filtering,
grouping, and plotting without any simulator-specific tooling.

\section{How to use the framework} \label{sec:how-to-use-sim}

The simulator is distributed as a Python package on Github\footnote{\url{https://github.com/MatthijsReyers/lora-simulator}}
and requires no external simulation framework and uses only very common dependencies like \textbf{asyncio} and \textbf{cffi}.
After cloning the package, a minimal simulation can be set up in a single script.
This section walks through a simple example: two LoRaWAN devices that exchange ping--pong messages through a gateway and network server, demonstrating OTAA activation, custom application handlers, and the simulation entry point.

\subsection{Defining application handlers}

Application-layer logic is implemented by subclassing the \texttt{Application} abstract base class and overriding two methods: \texttt{on\_uplink()} (called on the server side when a device sends data) and \texttt{on\_downlink()} (called on the device side when the server sends data back).
Each application registers itself on a specific FPort (1--223).

Listing~\ref{lst:ping-app} shows two small application classes.
The server-side \texttt{PongApp} listens for uplinks containing \texttt{b"ping"} and responds by queuing a \texttt{b"pong"} downlink.
The device-side \texttt{PingApp} simply logs any downlink it receives.

\begin{lstlisting}[caption={Server-side and device-side ping--pong application handlers.},label={lst:ping-app},float=htb]
from simulator.lorawan.application import Application
from simulator.lorawan.network_server import NetworkServer

class PongApp(Application):  # server-side
    def __init__(self, ns: NetworkServer, fport: int = 1):
        self.ns = ns
        self._fport = fport

    def port(self) -> int:
        return self._fport

    async def on_uplink(self, dev_addr: int, payload: bytes):
        if payload == b"ping":
            self.ns.queue_downlink(
                dev_addr, fport=self._fport, payload=b"pong")

    async def on_downlink(self, payload: bytes):
        pass  # server-side, not used

class PingApp(Application):  # device-side
    def port(self) -> int:
        return 1

    async def on_uplink(self, dev_addr: int, payload: bytes):
        pass  # device-side, not used

    async def on_downlink(self, payload: bytes):
        print(f"Device received: {payload}")
\end{lstlisting}

\subsection{Defining a device}

A device is created by subclassing \texttt{LoRaWanDevice}.
The constructor registers the device-side application and spawns a coroutine that performs the OTAA join procedure and then periodically transmits uplinks.
Listing~\ref{lst:ping-device} shows a device that sends a \texttt{b"ping"} uplink every 30 seconds.

\begin{lstlisting}[caption={A LoRaWAN device that sends periodic ping uplinks.},label={lst:ping-device},float=htb]
from simulator.environment import simulation_env as sim
from simulator.lorawan.device import LoRaWanDevice
from simulator.lorawan.join import OTAACredentials

class PingDevice(LoRaWanDevice):
    def __init__(self, otaa: OTAACredentials):
        super().__init__(otaa_credentials=otaa)
        self.register_application(PingApp())
        sim.create_task(self._ping_loop())
        self.device_class = 'C'  # enable continuous receive window for downlinks

    async def _ping_loop(self):
        await sim.sleep(1)  # let gateway start
        await self.join()   # OTAA handshake
        while sim.is_running():
            await self.send_uplink(fport=1, payload=b"ping")
            await sim.sleep(30)
\end{lstlisting}

\subsection{Running the simulation}

Finally, the simulation entry point instantiates the PHY layer, a network server and gateway, registers OTAA credentials, and starts the simulation loop.
Listing~\ref{lst:ping-main} shows the full setup.
Calling \texttt{sim.run()} advances the discrete-event loop until the specified simulation length (in seconds) is reached.

\begin{lstlisting}[caption={Simulation entry point for the ping--pong example.},label={lst:ping-main},float=htb]
from simulator.lora.phy_layer import LoraPhyLayer
from simulator.lora.radio import LoraRadio
from simulator.lorawan.gateway import LoRaWanGateway
from simulator.lorawan.network_server import NetworkServer
from simulator.lorawan.join import OTAACredentials

APP_EUI = bytes.fromhex("0102030405060708")
DEV_EUI = bytes.fromhex("1112131415161718")
APP_KEY = bytes.fromhex("2B7E151628AED2A6ABF7158809CF4F3C")

phy = LoraPhyLayer()  # shared wireless medium
ns  = NetworkServer()
ns.register_otaa_device(APP_EUI, DEV_EUI, APP_KEY)
ns.register_application(PongApp(ns))

gw = LoRaWanGateway(radio=LoraRadio(), network_server=ns)

creds = OTAACredentials(
    app_eui=APP_EUI, dev_eui=DEV_EUI, app_key=APP_KEY)
device = PingDevice(otaa=creds)

from simulator.environment import simulation_env as sim
sim.run(simulation_length=120)  # 2 minutes
\end{lstlisting}

Running this script produces log output showing the OTAA join handshake, the periodic \texttt{b"ping"} uplinks forwarded through the gateway, and the \texttt{b"pong"} downlinks delivered back to the device.
More examples---including custom non-LoRaWAN protocols, multicast groups, and containerized firmware setups---are available in the \texttt{examples/} directory of the repository.

\section{Conclusion} \label{sec:conclusion}

This paper presented the design of an open-source, Python-based LoRa
and LoRaWAN simulator that combines a physical-layer
model with a LoRaWAN~1.0.4 protocol stack and a containerized firmware
system for executing real STM32 C firmware inside the simulation.

The simulator is designed for ease of use (standard Python package, no
external simulation framework), modularity (replaceable path-loss,
collision, and protocol models), and extensibility (custom protocols
can be implemented without modifying the simulator core).
Its firmware-in-the-loop capability bridges the gap between simulation
and real-world deployment by allowing the same C source code to be
tested in both environments.

The simulator source code is publicly available at:
\url{https://github.com/MatthijsReyers/lora-simulator}.

\section{Acknowledgements}
We acknowledge the EU Project ENACT (https://enacthe.
eu/), which has received funding from the European Union
under the grant agreement No GA 101157151. Views and
opinions expressed are, however, those of authors only and do not necessarily reflect those of the European Union. Neither the European Union nor the granting authority can be held responsible for them.

\bibliographystyle{ieeetr}
\bibliography{references}

\end{document}